\documentclass[12pt]{iopart}
\usepackage{graphicx}

\begin{document}

\title[Quantum Correlations
of Two-Mode Gaussian Systems
]{Quantum Correlations
of Two-Mode Gaussian Systems
in a Thermal Environment}

\author{Aurelian Isar}

\address{National Institute of Physics and Nuclear Engineering,
P.O.Box MG-6, Bucharest-Magurele, Romania}
\ead{isar@theory.nipne.ro}
\begin{abstract}

In the framework of the theory of open systems based on completely positive quantum dynamical semigroups, we give a description of continuous variable quantum entanglement and quantum discord for a system consisting of two non-interacting non-resonant bosonic modes embedded in a thermal environment. We study the time evolution of logarithmic negativity, which characterizes the degree of entanglement, and show that in the case of an entangled initial squeezed thermal state, entanglement suppression takes place for all temperatures of the environment, including zero temperature. We analyze the time evolution of the Gaussian quantum discord, which is a measure of all quantum correlations in the bipartite state, including entanglement, and show that discord decays asymptotically in time under the effect of the thermal bath. We describe also the time evolution of classical correlations and quantum mutual information, which measures the total correlations of the quantum system.

\end{abstract}

\pacs{03.65.Yz, 03.67.Bg, 03.67.Mn}

\section{Introduction}

In recent years there is an increasing interest in using non-classical entangled states of continuous variable systems in applications of quantum information processing, communication and computation \cite{bra1}. In this respect, Gaussian states, in particular two-mode Gaussian states, play a key role since they can be easily created and controlled experimentally. Due to the unavoidable interaction with the environment, in order to describe realistically quantum information processes it is necessary to take decoherence and dissipation into consideration. Decoherence and dynamics of quantum entanglement in continuous variable open systems have been intensively studied in the last years \cite{oli,ser3,pra,dod1,ser4,avd,ben1,mch,man,jan,aphysa,aeur,arus1,paz2,arus2,ascri1}.

In this paper we study, in the framework of the theory of open systems based on completely positive quantum dynamical semigroups, the dynamics of continuous variable quantum entanglement and quantum discord of a subsystem consisting of two uncoupled bosonic modes (harmonic oscillators) interacting with a common thermal environment. We are interested in discussing the correlation effect of the environment, therefore we assume that the two modes are independent, i.e. they do not interact directly. The initial state of the subsystem is taken of Gaussian form and the evolution under the quantum dynamical semigroup assures the preservation in time of the Gaussian form of the state. In Refs. \cite{ascri4,aosid1,ascri2} we studied the evolution of entanglement and quantum discord of two identical harmonic oscillators interacting with a thermal environment for initial symmetric squeezed vacuum and squeezed thermal states of the subsystem. In this work we extend the previous analysis to the case of non-resonant bosonic modes and take non-symmetric squeezed thermal states as initial states. We show that in the case of an entangled initial squeezed thermal state, entanglement suppression (entanglement sudden death) takes place for all temperatures of the environment, including zero temperature. We analyze the time evolution of Gaussian quantum discord, which is a measure of all quantum correlations in the bipartite state, including entanglement, and show that discord decays asymptotically in time under the effect of the thermal bath. Before the suppression of the entanglement, the qualitative evolution of discord is very similar to that of the entanglement. We describe also the time evolution of classical correlations and quantum mutual information, which measures the total correlations of quantum system.

\section{Equations of motion for two modes interacting with an environment}

We study the dynamics of a subsystem composed of two non-interacting bosonic modes in weak interaction with a thermal environment. In the axiomatic formalism based on completely positive quantum dynamical semigroups, the Markovian irreversible time evolution of an open system is described by the Kossakowski-Lindblad master equation \cite{rev,san}.
We are interested in the set of Gaussian states, therefore we introduce such quantum dynamical semigroups that preserve this set during time evolution of the system. The Hamiltonian of the two uncoupled non-resonant harmonic oscillators of identical mass $m$ and frequencies $\omega_1$ and $\omega_2$ is
\begin{eqnarray} H={1\over 2m}(p_x^2+p_y^2)+\frac{m}{2}(\omega_1^2 x^2+\omega_2^2 y^2),\end{eqnarray} where $x,y$ are the coordinates and $p_x,p_y$ are the momenta of the two quantum oscillators.

The equations of motion for the quantum correlations of the canonical observables $x,y$ and $p_x,p_y$ are the following ($\rm T$ denotes the transposed matrix) \cite{san}:
\begin{eqnarray}{d \sigma(t)\over
dt} = Y \sigma(t) + \sigma(t) Y^{\rm T}+2 D,\label{vareq}\end{eqnarray} where
\begin{equation} Y=\left(\matrix{ -\lambda&1/m&0 &0\cr -m\omega_1^2&-\lambda&0&
0\cr 0&0&-\lambda&1/m \cr 0&0&-m\omega_2^2&-\lambda}\right),~~
D=\left(\matrix{
D_{xx}& D_{xp_x} &D_{xy}& D_{xp_y} \cr D_{xp_x}&D_{p_x p_x}&
D_{yp_x}&D_{p_x p_y} \cr D_{xy}& D_{y p_x}&D_{yy}& D_{y p_y}
\cr D_{xp_y} &D_{p_x p_y}& D_{yp_y} &D_{p_y p_y}} \right),\end{equation}
and the diffusion coefficients $D_{xx}, D_{xp_x},$... and the dissipation constant $\lambda$ are real quantities. We introduced the following $4\times 4$ bimodal covariance matrix:
\begin{eqnarray}\sigma(t)=\left(\matrix{\sigma_{xx}(t)&\sigma_{xp_x}(t) &\sigma_{xy}(t)&
\sigma_{xp_y}(t)\cr \sigma_{xp_x}(t)&\sigma_{p_xp_x}(t)&\sigma_{yp_x}(t)
&\sigma_{p_xp_y}(t)\cr \sigma_{xy}(t)&\sigma_{yp_x}(t)&\sigma_{yy}(t)
&\sigma_{yp_y}(t)\cr \sigma_{xp_y}(t)&\sigma_{p_xp_y}(t)&\sigma_{yp_y}(t)
&\sigma_{p_yp_y}(t)}\right)=\left(\begin{array}{cc}A&C\\
C^{\rm T}&B \end{array}\right),\label{covar}
\end{eqnarray}
where $A$, $B$ and $C$ are $2\times 2$ Hermitian matrices. $A$ and $B$ denote the symmetric covariance matrices for the individual reduced one-mode states, while the matrix $C$ contains the cross-correlations between modes.
The elements of the covariance matrix are defined as $\sigma_{ij}=<R_iR_j+R_jR_i>/2, i,j=1,..,4,$ with ${\bf R}=\{x,p_x,y,p_y\},$ which up to local displacements fully characterize any Gaussian state of a bipartite system.

The time-dependent solution of Eq. (\ref{vareq}) is given by \cite{san}
\begin{eqnarray}\sigma(t)= M(t)[\sigma(0)-\sigma(\infty)] M^{\rm T}(t)+\sigma(\infty),\label{covart}\end{eqnarray} where the matrix $M(t)=\exp(Yt)$ has to fulfill the condition $\lim_{t\to\infty} M(t) = 0.$ The values at infinity are obtained from the equation \begin{eqnarray}
Y\sigma(\infty)+\sigma(\infty) Y^{\rm T}=-2 D.\label{covarinf}\end{eqnarray}

\section{Dynamics of quantum correlations}

\subsection{Time evolution of entanglement}

 In order to quantify the degree of entanglement of the two-mode states it is appropriate to use the logarithmic negativity. For a Gaussian density operator, the logarithmic negativity is completely defined by the symplectic spectrum of the partial transpose of the covariance matrix. It is given by
$E_N={\rm max}\{0,-\log_2 2\tilde\nu_-\},$
where $\tilde\nu_-$ is the smallest of the two symplectic eigenvalues of the partial transpose $\tilde{{\sigma}}$ of the two-mode covariance matrix $\sigma$ \cite{ser4}:
\begin{eqnarray}2\tilde{\nu}_{\mp}^2 = \tilde{\Delta}\mp\sqrt{\tilde{\Delta}^2
-4\det\sigma}
\end{eqnarray}
and $ \tilde\Delta$ is the symplectic invariant (seralian), given by
$ \tilde\Delta=\det A+\det B-2\det C.$

In our model, the logarithmic negativity is calculated as \cite{aijqi,aosid} \begin{eqnarray}E_N(t)={\rm max}\{0,-\frac{1}{2}\log_2[4g(\sigma(t))]\}, \end{eqnarray} where \begin{eqnarray}g(\sigma(t))=\frac{1}{2}(\det A +\det
B)-\det C\nonumber\\
-\left({\left[\frac{1}{2}(\det A+\det B)-\det
C\right]^2-\det\sigma(t)}\right)^{1/2}.\end{eqnarray}
It determines the strength of entanglement for $E_N(t)>0,$ and if $E_N(t)=0,$ then the state is separable.

We assume that the initial Gaussian state is a two-mode squeezed thermal state, with the covariance matrix of the form \cite{mar}
\begin{eqnarray}\sigma_{st}(0)=\frac{1}{2}\left(\matrix{a&0&c&0\cr
0&a&0&-c\cr
c&0&b&0\cr
0&-c&0&b}\right),\label{ini1} \end{eqnarray}
with the matrix elements given by
\begin{eqnarray}a=n_1 \cosh^2 r + n_2 \sinh^2 r + \frac{1}{2} \cosh 2r,\\
b=n_1 \sinh^2 r + n_2 \cosh^2 r + \frac{1}{2} \cosh 2r,\\
c=\frac{1}{2}(n_1 + n_2 + 1) \sinh 2r,\label{ini2}
\end{eqnarray}
where $n_1,n_2$ are the average number of thermal photons associated with the two modes and $r$ denotes the squeezing parameter.
In the particular case $n_1=0$ and $n_2=0$, (\ref{ini1}) becomes the covariance matrix of the two-mode squeezed vacuum state. A two-mode squeezed thermal state is entangled when the squeezing parameter $r$ satisfies the inequality $r>r_s$ \cite{mar},
where \begin{eqnarray} \cosh^2 r_s=\frac{(n_1+1)(n_2+1)}{ n_1+n_2+1}. \end{eqnarray}

We suppose that the asymptotic state of the considered open system is a Gibbs state corresponding to two independent bosonic modes in thermal equilibrium at temperature $T.$ Then the quantum diffusion coefficients have the following form (we put from now on $\hbar=1$) \cite{rev}:
\begin{eqnarray}m\omega_1 D_{xx}=\frac{D_{p_xp_x}}{m\omega_1}=\frac{\lambda}{2}\coth\frac{\omega_1}{2kT},\nonumber\\
m\omega_2 D_{yy}=\frac{D_{p_yp_y}}{m\omega_2}=\frac{\lambda}{2}\coth\frac{\omega_2}{2kT},\label{envcoe}\\
D_{xp_x}=D_{yp_y}=D_{xy}=D_{p_xp_y}=D_{xp_y}=D_{yp_x}=0.\nonumber\end{eqnarray}

The evolution of entangled initial squeezed thermal states with the covariance matrix given by Eq. (\ref{ini1}) is illustrated in Fig. 1, where we represent the dependence of the logarithmic negativity $E_N(t)$ on time $t$ and temperature $T$ for the case of an initial non-symmetric Gaussian state ($a\neq b$). For all temperatures $T,$ including zero temperature, at certain finite moment of time, which depends on $T,$ $E_N(t)$ becomes zero and therefore the state becomes separable. This is the so-called phenomenon of entanglement sudden death. It is in contrast to the quantum decoherence, during which the loss of quantum coherence is usually gradual \cite{aphysa,arus}. One can also show that the dissipation favors the phenomenon of entanglement sudden death -- with increasing the dissipation parameter $\lambda,$ the entanglement suppression happens earlier. The same qualitative behaviour of the time evolution of entanglement was obtained previously \cite{aosid1,ascri3} in the particular case $n_1=0$ and $n_2=0$ corresponding to an initial two-mode squeezed vacuum state and in the case of symmetric initial squeezed thermal states.

Comparing the present results with those obtained in the previous paper \cite{ascri2} one can assert that the asymmetry ($a\neq b$) of the initial Gaussian state favors the suppression of entanglement. The most robust under the influence of the environment is the entanglement of symmetric ($a=b$) initial squeezed thermal states. An even stronger influence on the entanglement has the non-resonant character of the two modes: by increasing the ratio of the frequencies of the two modes, the entanglement sudden death happens earlier in time. The longest surviving entanglement takes place when the modes are resonant ($\omega_1=\omega_2$). This effect due to the non-resonance of the modes is stronger for small values of the frequencies, and it diminishes, for the same ratio of frequencies, by increasing the values of frequencies.

In our model, in which we suppose that the asymptotic state of the considered open system is a Gibbs state corresponding to two independent bosonic modes in thermal equilibrium, an separable initial state remains separable in time, and it is not possible to generate entanglement. This is in contrast with the possibility of entanglement generation starting, for instance, with a separable state in the case of two non-interacting two-level systems immersed in a common bath \cite{ben2}. At the same time we remind that we have studied previously \cite{aosid,ascri} the evolution of the entanglement of two identical harmonic oscillators interacting with a general environment, characterized by general diffusion and dissipation coefficients, and we obtained that, for separable initial states and for definite values of these coefficients, entanglement generation or a periodic generation and collapse of entanglement take place. In discussing the entanglement decay, it is interesting to mention that models have been elaborated to realize quantum feedback control of continuous variable entanglement for a system consisting of two interacting bosonic modes plunged into an environment, based on a local technique \cite{man1}, or on a nonlocal homodyne measurement \cite{man2}.

\begin{figure}
\resizebox{0.4\columnwidth}{!}
{
\includegraphics{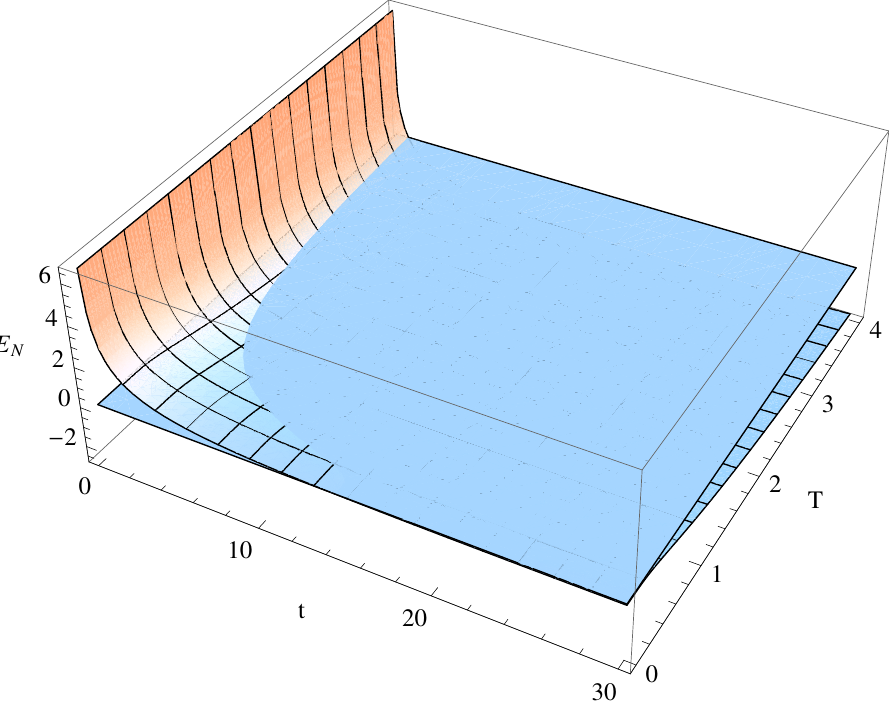}
}
\caption{Logarithmic negativity $E_N$ versus time $t$ and temperature $T$ for an entangled initial non-symmetric squeezed thermal state with squeezing parameter $r=3$, $n_1=3, n_2=1$ and $\lambda=0.1, \omega_1=1, \omega_2=2.$ We take $m=\hbar=k=1.$
}
\label{fig:1}
\end{figure}

\subsection{Gaussian quantum discord}

Recent studies have shown that separable states, usually considered as being classically correlated, might also contain quantum correlations. Quantum discord was introduced \cite{zur,oll} as a measure of all quantum correlations in a bipartite state, including -- but not restricted to -- entanglement. Quantum discord has been defined as the difference
between two quantum analogues of classically equivalent expressions of the mutual information, which is a measure of total correlations in a quantum state.
For an arbitrary bipartite state $\rho_{12},$ the total correlations are expressed by quantum mutual information
\begin{eqnarray}
I(\rho_{12})=\sum_{i=1,2} S(\rho_{i})-S(\rho_{12}),
\end{eqnarray} where $\rho_i$ represents the reduced density matrix of subsystem $i$ and $S(\rho)= - {\rm Tr}(\rho \ln \rho)$
is the von Neumann entropy. A measure of bipartite classical correlations $C(\rho_{12})$ in the bipartite quantum state $\rho_{12}$ based on a complete set of local projectors $\{\Pi_{2}^k\}$ on the subsystem 2 can be given by
\begin{eqnarray}
C(\rho_{12})=S(\rho_{1})-{\inf}_{\{\Pi_{2}^k\}}\{S(\rho_{1|2})\},
\end{eqnarray}
where $S(\rho_{1|2}) =\sum_{k}p^k S(\rho_{1}^k)$ is the conditional entropy of subsystem {1} and $\inf\{S(\rho_{1|2})\}$
represents the minimal value of the entropy with respect to a complete set of local measurements $\{\Pi_{2}^k\}.$
Here, $p^k$ is the measurement probability for the $k$th local projector and $\rho_{1}^k$ denotes the reduced state of subsystem $1$ after the local measurements.
Then the quantum discord is defined by
\begin{eqnarray}
D(\rho_{12})=I(\rho_{12})-C(\rho_{12}).
\end{eqnarray}

Originally the quantum discord was defined and evaluated mainly for finite dimensional systems. Recently \cite{par,ade} the notion of discord has been extended to the domain of
continuous variable systems, in particular to the analysis of bipartite systems described by two-mode Gaussian states.
Closed formulas have been derived for bipartite thermal squeezed states \cite{par} and for all two-mode Gaussian states \cite{ade}.

The Gaussian quantum discord of a general two-mode Gaussian state $\rho_{12}$ can be defined as the quantum discord where the conditional entropy is restricted to generalized Gaussian positive operator valued measurements (POVM) on the mode 2, and in terms of symplectic invariants it is given by (the symmetry between the two modes 1 and 2 is broken) \cite{ade}
\begin{eqnarray}
D=f(\sqrt{\beta})-f(\nu_-) - f(\nu_+) + f(\sqrt{\varepsilon}),
\label{disc}
\end{eqnarray}
where \begin{eqnarray}f(x) =\frac{x+1}{2} \log\frac{x+1}{2} -\frac{x-1}{2} \log\frac{x-1}{2},\end{eqnarray}
\begin{eqnarray}\label{infdet}
\varepsilon=
 &  &
\hspace*{-.1cm}
\left\{  \hspace*{-.5cm}  \begin{array}{rcl}& &\begin{array}{c}\displaystyle{\frac{{2 \gamma^2+(\beta-1)(\delta-\alpha)
+2 |\gamma| \sqrt{\gamma^2+(\beta-1) (\delta-\alpha)}}}{{(\beta-1){}^2}}}\end{array},\\& &\qquad
\hbox{if}~~(\delta-\alpha\beta)^2 \le (\beta+1)\gamma^2 (\alpha +\delta)\\ \\& &
\begin{array}{c}\displaystyle{\frac{{\alpha\beta-\gamma^2+\delta-\sqrt{\gamma^4+(\delta-\alpha\beta){}^2-
2\gamma^2(\delta+\alpha\beta)}}}{{2\beta}}}\end{array}, \\& & \qquad
\hbox{otherwise,} \end{array} \right.
\end{eqnarray}
\begin{eqnarray}\alpha=4\det A,~~~\beta=4\det B,~~~\gamma=4\det C,~~~\delta=16\det\sigma,\end{eqnarray}
and $\nu_\mp$ are the symplectic eigenvalues of the state, given by
\begin{eqnarray}2{\nu}_{\mp}^2 ={\Delta}\mp\sqrt{{\Delta}^2
-4\det\sigma},
\end{eqnarray}
where
$\Delta=\det A+\det B+2\det C.$

\begin{figure}
\resizebox{0.4\columnwidth}{!}
{
\includegraphics{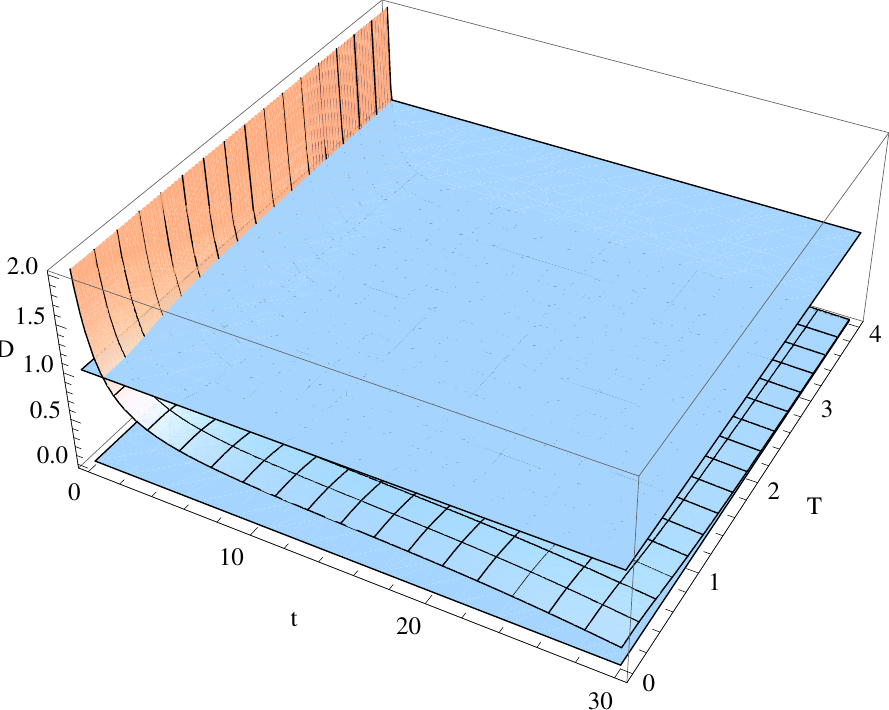}
}
\caption{Gaussian quantum discord $D$ versus time $t$ and temperature $T$ for an entangled initial non-symmetric squeezed thermal state with squeezing parameter $r=3$, $n_1=3, n_2=1$ and $\lambda=0.1, \omega_1=1, \omega_2=2.$ We take $m=\hbar=k=1.$
}
\label{fig:2}
\end{figure}

The evolution of the Gaussian quantum discord $D$ is illustrated in Figure 2, where we represent the dependence of $D$ on time $t$ and temperature $T$ for an entangled initial non-symmetric Gaussian state, taken of the form of a two-mode squeezed thermal state (\ref{ini1}), for such values of the parameters which satisfy for all times the first condition in formula (\ref{infdet}). The Gaussian discord has nonzero values for all finite times and this fact certifies the existence of non-classical correlations in two-mode Gaussian states, either separable or entangled. Gaussian discord asymptotically decreases in time, compared to the case of logarithmic negativity, which has an evolution leading to a sudden suppression of entanglement. For entangled initial states the Gaussian discord remains strictly positive in time and in the limit of infinite time it tends asymptotically to zero, corresponding to the thermal product (separable) state, with no correlations at all.

From Fig. 2 we notice that, in agreement with the general properties of the Gaussian quantum discord \cite{ade}, the states can be either separable or entangled for $D\le 1$ and all the states above the threshold $D=1$ are entangled. We also notice that the decay of quantum discord is stronger when the temperature $T$ is increasing.
It should be remarked that the decay of quantum discord is very similar to that of the entanglement before the time of the sudden death of entanglement. Near the threshold of zero logarithmic negativity ($E_N = 0$), the nonzero values of the discord can quantify the non-classical correlations for separable mixed states and one considers that this fact could make possible some tasks in quantum computation \cite{yut}.
The discord is increasing with the squeezing parameter $r$ and it is decreasing with increasing the ratio of the frequencies $\omega_1$ and $\omega_2$ of the two modes and the difference of parameters $a$ and $b.$

\begin{figure}
\resizebox{0.4\columnwidth}{!}
{
\includegraphics{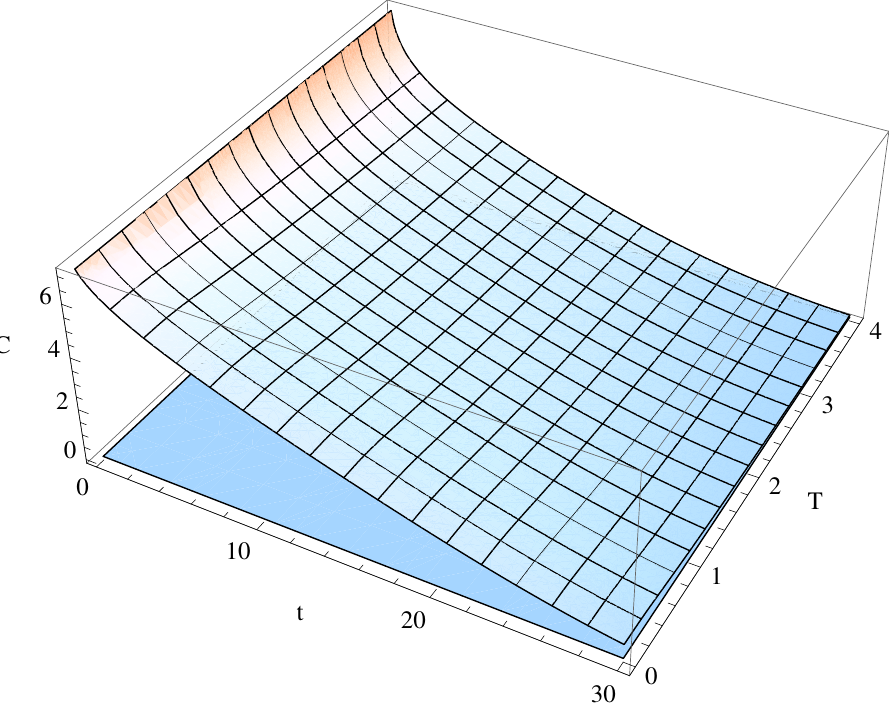}
}
\caption{Degree of classical correlations $C$ versus time $t$ and temperature $T$ for an entangled initial non-symmetric squeezed thermal state with squeezing parameter $r=3$, $n_1=3, n_2=1$ and $\lambda=0.1, \omega_1=1, \omega_2=2.$ We take $m=\hbar=k=1.$
}
\label{fig:3}
\end{figure}

\subsection{Classical corellations and quantum mutual information}

The measure of classical correlations for a general two-mode Gaussian state $\rho_{12}$ can also be calculated and it is given by \cite{ade}
\begin{eqnarray}
C=f(\sqrt{\alpha}) - f(\sqrt{\varepsilon}),
\label{clas}
\end{eqnarray}
while the expression of the quantum mutual information, which measures the total correlations, is given by
\begin{eqnarray}
I=f(\sqrt{\alpha}) + f(\sqrt{\beta}) -f(\nu_-) - f(\nu_+).
\label{mut}
\end{eqnarray}

In Figs. 3 and 4 we illustrate the evolution of classical correlations $C$ and, respectively, quantum mutual information $I,$ as functions of time $t$ and temperature $T$ for an entangled initial Gaussian state, taken of the form of a two-mode squeezed thermal state (\ref{ini1}). These two quantities manifest a qualitative behaviour similar to that one of the Gaussian discord: they have nonzero values for all finite times and in the limit of infinite time they tend asymptotically to zero, corresponding to the thermal product (separable) state, with no correlations at all. One can also see that the classical correlations and quantum mutual information decrease with increasing the temperature of the thermal bath.
One can show that the classical correlations and quantum mutual information increase with increasing the squeezing parameter $r$ and the difference of parameters $a$ and $b.$ At the same time classical correlations increase with the ratio of the frequencies $\omega_1$ and $\omega_2$ of the two modes, while quantum mutual information is decreasing with increasing this ratio.

For comparison all these quantities are represented also on the same graphic in Fig. 4.  In the considered case the value of classical correlations is larger than that of quantum correlations, represented by the Gaussian quantum discord.

\begin{figure}
\resizebox{0.4\columnwidth}{!}
{
\includegraphics{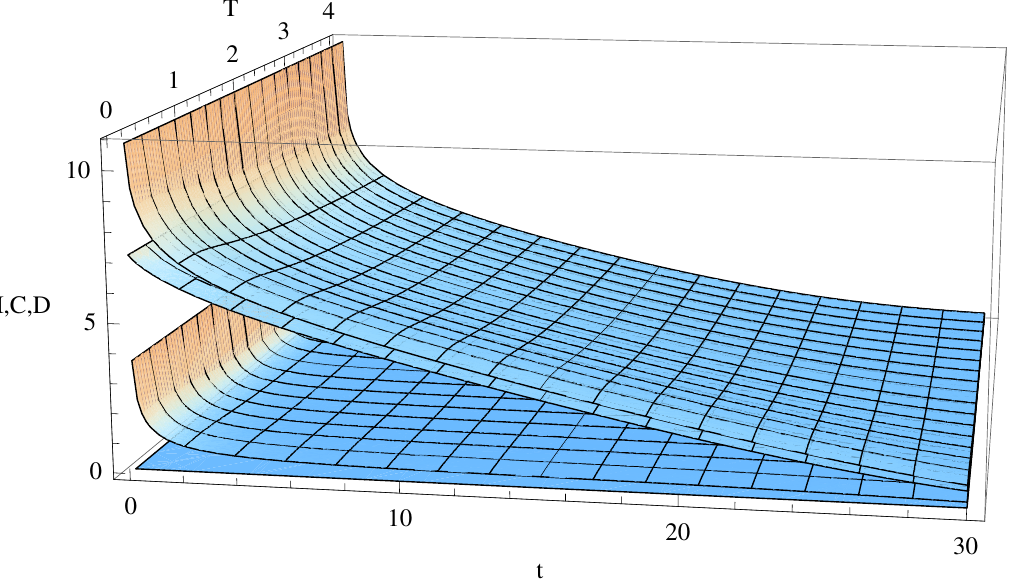}
}
\caption{Quantum mutual information $I$ versus time $t$ and temperature $T$ for an entangled initial non-symmetric squeezed thermal state with squeezing parameter $r=3$, $n_1=3, n_2=1$ and $\lambda=0.1, \omega_1=1, \omega_2=2.$ We take $m=\hbar=k=1.$ There are also represented the Gaussian quantum discord and classical correlations.
}
\label{fig:4}
\end{figure}

\section{Summary}

We investigated the Markovian dynamics of quantum correlations for a subsystem composed of two non-interacting bosonic modes embedded in a thermal bath. We have analyzed the influence of the environment on the dynamics of quantum entanglement and quantum discord for Gaussian initial states. We have described the time evolution of the logarithmic negativity in terms of the covariance matrix for non-symmetric squeezed thermal states, for the case when the asymptotic state of the considered open system is a Gibbs state corresponding to two independent quantum harmonic oscillators in thermal equilibrium. The dynamics of the quantum entanglement strongly depends on the initial states and the parameters characterizing the environment (dissipation coefficient and temperature). For an entangled initial squeezed thermal state, entanglement suppression (entanglement sudden death) takes place for all values of the temperatures of the environment, including zero temperature. The time when the entanglement is suppressed decreases with increasing the temperature and dissipation.

We described also the time evolution of Gaussian quantum discord, which is a measure of all quantum correlations in the bipartite state, including entanglement.
The values of quantum discord decrease asymptotically in time. This is in contrast to the sudden death of entanglement. The time evolution of quantum discord is very similar to that of entanglement before the sudden suppression of the entanglement. Quantum discord is decreasing with increasing the temperature. After the sudden death of entanglement the nonzero values of discord manifest the existence of quantum correlations for separable mixed states. We described also the time evolution of classical correlations and quantum mutual information, which measures the total correlations of the quantum system.

\ack

The author thanks the referee for his useful suggestions and recommendations. The author acknowledges the financial support received from the Romanian Ministry of Education and Research, through the Projects CNCS-UEFISCDI PN-II-ID-PCE-2011-3-0083 and PN 09 37 01 02/2010.

\end{document}